# Fortuitous partners of antiferromagnetic and Mott states in spin-orbit-coupled $Sr_2IrO_4$: A study of $Sr_2Ir_{1-x}M_xO_4$ (M=Fe or Co)

Bing Hu[1,2], Hengdi Zhao[1], Yu Zhang[1], Pedro Schlottmann[3], Feng Ye[4] and Gang Cao[1*]

[1]Department of Physics, University of Colorado at Boulder, Boulder, Colorado 80309, USA

[2]School of Mathematics and Physics, North China Electric Power University, Beijing 102206, China

[3]Department of Physics, Florida State University, Tallahassee, Florida 32306, USA

[4] Neutron Scattering Division, Oak Ridge National Laboratory, Oak Ridge, TN 37831, USA

$Sr_2IrO_4$ is an archetypal spin-orbit-coupled Mott insulator with an antiferromagnetic state below 240 K. Here we report results of our study on single crystals of $Sr_2Ir_{1-x}\mathbf{Fe}_xO_4$ ($0 \leq x < 0.32$) and $Sr_2Ir_{1-x}\mathbf{Co}_xO_4$ ($0 \leq x < 0.22$). Fe doping retains the antiferromagnetic state but simultaneously precipitates an emergent metallic state whereas Co doping causes a rapid collapse of both the antiferromagnetic and Mott states, giving rise to a confined metallic state featuring a pronounced linearity of the basal-plane resistivity up to 700 K. The results indicate tetravalent $Fe^{4+}(3d^4)$ ions in the intermediate spin state with S=1 and $Co^{4+}(3d^5)$ ions in the high spin state with S=5/2 substituting for $Ir^{4+}(5d^5)$ ions in $Sr_2IrO_4$, respectively. The effective magnetic moment closely tracks the Néel temperature as doping increases, suggesting that the spin state of the dopant predominately determines the magnetic properties in doped $Sr_2IrO_4$. Furthermore, all relevant properties including charge-carrier density (e.g., $10^{28}/m^3$), Sommerfeld coefficient (e.g., 19 mJ/mole $K^2$) and Wilson ratio (e.g., 2.6), consistently demonstrates a metallic state that is both robust and highly correlated in the two systems, arising from the percolation of bound states and the weakening of structural distortions. This study strongly suggests that the antiferromagnetic and Mott states merely coexist in a fortuitous manner in $Sr_2IrO_4$.

*gang.cao@colorado.edu

**I. Introduction**

The spin-orbit-coupled Mott state in $Sr_2IrO_4$ is an intriguing consequence of a delicate interplay of on-site Coulomb repulsion, U, strong spin-orbit interactions (SOI), and crystalline field effects [1-8]. The iridate is among the most extensively studied quantum materials in recent years in part because of its apparent structural and magnetic similarities to those of the cuprate $La_2CuO_4$ and the widely anticipated novel superconductivity in electron-doped $Sr_2IrO_4$ [9-14], which, however, has remained elusive thus far. This conspicuous absence of the anticipated superconductivity brings to light a novel and yet underappreciated nature of the spin-orbit-coupled Mott state, making $Sr_2IrO_4$ even more interesting and extraordinary for further investigations.

$Sr_2IrO_4$ adopts a canted antiferromagnetic (AFM) state [15] with a Néel temperature $T_N$ = 240 K [16-19] and an energy gap $\Delta \leq 0.62$ eV [20-22]. Empirical trends indicate that the iridate lacks a conventional correlation between the magnetic and insulating states, that is, the AFM order does not always track changes of the insulating state in $Sr_2IrO_4$ [5]. Recent studies show that the insulating state in $Sr_2IrO_4$ unusually persists at megabar pressures up to 185 GPa [23], and yet the AFM state already diminishes at 17 GPa [24], which inspires expectations of pressured-induced quantum paramagnetism [25] and an exotic insulating state [23]. In contrast, the ground state of the iridate is far more susceptible to impurity doping. Oftentimes, a slight substitution at either the $Sr^{2+}$ site (e.g., $La^{3+}$, $Ca^{2+}$ and $Ba^{2+}$) [26-28] or the $Ir^{4+}$ site (e.g., $Mn^{3+}$, $Ru^{4+}$, $Rh^{3+}$ and $Tb^{4+}$) [29-39] causes disproportionately large changes in either the insulating state or the magnetic state or both. However, each dopant mentioned above generates unique, quite often exotic behavior in its own right, implying a rich phase diagram of the spin-orbit-coupled Mott state that is yet to be fully explored and understood.

In this paper, we report results of our study on single crystals of Fe and Co doped $Sr_2IrO_4$ or $Sr_2Ir_{1-x}$**$Fe_x$**$O_4$ ($0 \leq x < 0.32$) and $Sr_2Ir_{1-x}$**$Co_x$**$O_4$ ($0 \leq x < 0.22$). The key observations are that Fe doping preserves the AFM state but simultaneously precipitates an emergent metallic state; in contrast, Co doping causes a rapid collapse of both the AFM and Mott states, giving rise to a highly anisotropic or confined metallic state that features a pronounced linearity of the basal-plane resistivity extending up to 700 K with no sign of saturation. The detailed experimental results including structural, transport, thermal and magnetic properties indicate that it is tetravalent $Fe^{4+}$(*$3d^4$*) and $Co^{4+}$(*$3d^5$*) ions that substitute tetravalent $Ir^{4+}$(*$5d^5$*) ions in $Sr_2IrO_4$, respectively. The crystalline field of the octahedra splits the *3d* states into a *$t_{2g}$* ground-triplet and an excited *$e_g$* doublet. Hence, the crystalline field splitting quenches the angular momentum and competes with the Hund exchange that tends to maximize the spin in the *3d*-ion. Three spin states are possible for both, $Fe^{4+}$ and $Co^{4+}$, namely a high-spin state (HS), an intermediate spin state (IS), and a low-spin state (LS). For $Fe^{4+}$ these states have spin $S=2$, $S=1$ and $S=0$, respectively, while for $Co^{4+}$ the spins are $S=5/2$, $3/2$ and ½. Since Hund's rules are local interactions, the underlying SOI environment in the iridates plays only a secondary role for the impurities. The effective moment of the compounds grows with increasing doping, for both $Fe^{4+}$ and $Co^{4+}$. The Néel temperature in the Fe-compound is not affected up to $x=0.2$, but $T_N$ drops rapidly with increasing Co substitution. All the results indicate that the spin state of the dopant predominately determines the magnetic properties in doped $Sr_2IrO_4$.

Furthermore, an array of significantly enhanced characteristic parameters, such as charge-carrier density ($10^{27}$, $10^{28}$/$m^3$), Sommerfeld coefficient (30, 19 mJ/mole $K^2$) and Wilson ratio (2.7, 2.6), consistently demonstrates a metallic state that is both robust and highly correlated in $Sr_2Ir_{1-x}$**$Fe_x$**$O_4$ and $Sr_2Ir_{1-x}$**$Co_x$**$O_4$. In these materials, the $Fe^{4+}$(*$3d^4$*) ions offer additional holes leading to hydrogen-like acceptor states of large Bohr radius, which rapidly percolate into an impurity band with increasing x. On the other hand, the neutral substitution of $Co^{4+}$ ($3d^5$) ions adds neither holes nor electrons but breaks the translational invariance of the lattice leading to a small bound-state of the size of a unit cell [40]. Both impurities locally reduce the SOI effect of the undoped iridate [1] and significantly relax the structural distortions inherent in $Sr_2IrO_4$. This study underscores that the AFM and insulating states merely coexist in a fortuitous manner in $Sr_2IrO_4$, which sharply contrasts situations in other correlated materials such as $La_2CuO_4$, whose hallmark is a strong correlation between the AFM and Mott states. Note that synthesis and characterization of Fe and Co doped $Sr_2IrO_4$ in polycrystalline form were reported in Ref. [41]. However, our study has little overlap with Ref. [41] in terms of content and conclusion.

## II. Experimental

Single crystals of Fe or Co doped $Sr_2IrO_4$ were grown using a flux method. The starting materials were $SrCO_3$, $SrCl_2$, $IrO_2$ and $Fe_2O_3$ or $Co_3O_4$. The mixtures were fired in Pt crucibles at 1440 C for 15 hours and then slowly cooled to room temperature at a rate of 3 C/hour. Measurements of crystal structures were performed using a Bruker Quest ECO single-crystal diffractometer equipped with a PHOTON 50 CMOS detector. It is also equipped with an Oxford Cryosystem that creates sample temperature environments ranging from 80 K to 400 K during x-ray diffraction measurements. Chemical analyses of the samples were performed using a combination of a Hitachi MT3030 Plus Scanning Electron Microscope and an Oxford Energy Dispersive X-Ray Spectroscopy (EDX). Magnetic properties were measured using a Quantum Design (QD) MPMS-7 SQUID Magnetometer. The measurements of the electrical resistivity, Hall effect were carried out using a QD Dynacool PPMS System equipped with a 14-Tesla magnet and a dilution refrigerator. High temperature resistivity up to 700 K was measured using a home-made setup. The heat capacity was measured down to 0.05 K using a dilution refrigerator for the PPMS.

### III. Results and Discussion

### (a) Structural

The dopant Fe or Co substitutes Ir with a doping level up to 32% and 22%, respectively. The Fe or Co doping retains the native crystal structure but significantly alters the lattice parameters of $Sr_2Ir_{1-x}M_xO_4$ (M=Fe or Co), as shown in **Figs. 1**. A remarkable trend is that the unit cell volume V shrinks with increasing x; specifically, V is smaller by 0.46% and 0.48% at x=0.31 of Fe doping and x=0.17 of Co doping, respectively, compared to V for x=0 (**Fig.1f**). The reduction of V, along with a shortening of the basal-plane bond length Ir-O2 (**Figs.1b** and **1g**), indicates that the ionic radius of the dopant, Fe or Co, must be smaller than that of the $Ir^{4+}$ ($5d^5$) ion. Of all possible oxidation states of the Fe and Co ions, it is most likely that the $Fe^{4+}$ ($3d^4$) or $Co^{4+}$ ($3d^5$) ion substitutes the $Ir^{4+}$ ion in $Sr_2IrO_4$ because the ionic radius, $r$, of both $Fe^{4+}$ and $Co^{4+}$ ions is smaller than that of the $Ir^{4+}$ ion, i.e., $r(Ir^{4+}) = 0.625$ Å $> r(Fe^{4+}) = 0.585$ Å $> r(Co^{4+}) = 0.530$ Å. That $r(Co^{4+})$ is smaller than $r(Fe^{4+})$ explains the more rapid reduction in V with Co doping, consistent with the empirical Vegard's law (**Fig.1f**). This argument rules out possibilities of lower oxidation states of Fe and Co because due to the screened Coulomb potential of the nucleus the ionic radius is a strong function of the oxidation state, decreasing as $d$ electrons are removed, that is, the lower the oxidation state an ion has, the larger the ionic radius of the ion becomes. For example, $r(Fe^{3+}(3d^5)) = 0.645$ Å $> r(Fe^{4+}(3d^4))$, and $r(Co^{3+}(3d^6)) = 0.610$ Å $> r(Co^{4+}(3d^5))$. Note that $r(Co^{3+})$ is slightly smaller than $r(Ir^{4+})$ (=0.625 Å), but the rapid and significant volume reduction due to Co doping, e.g., $\Delta V/V=0.48$% at x=0.17 (compared to 0.46% at x=0.31 of Fe doping), makes the $Co^{3+}$ ion an unlikely replacement for $Ir^{4+}$ in $Sr_2IrO_4$. This point is also corroborated by the sizable shortening of the basal-plane bond length Ir-O2 (**Fig.1g**).

In short, the above structural analysis leads to the assignment of oxidation state of $Fe^{4+}$ ($3d^4$) and $Co^{4+}$ ($3d^5$). This means that an $Fe^{4+}$ ion with four $3d$-electrons provides hole doping whereas a $Co^{4+}$ ion with five $3d$-electrons offers no additional holes or electrons in $Sr_2IrO_4$.

Moreover, the chemical doping also lessens structural distortions inherent in $Sr_2IrO_4$ as the Ir-O-Ir bond angle (**Fig.1b**) relaxes with increasing x (**Fig.1h**), indicating that the Fe or Co dopant significantly reduces the rotation of octahedral $IrO_6$, which plays an important role in determining the physical properties [5,6,15,42,43].

**(b) Magnetic properties**

We now examine the magnetic properties of $Sr_2Ir_{1-x}Fe_xO_4$ and $Sr_2Ir_{1-x}Co_xO_4$ illustrated in **Fig.2**.

*$Sr_2Ir_{1-x}Fe_xO_4$*. One striking observation is that Fe doping up to 20% essentially retains the native Néel temperature $T_N$. Both the a-axis and c-axis magnetization, $M_a(T)$ and $M_c(T)$, exhibit little shift in $T_N$ for $x \leq 0.18$, as shown in **Figs.2a** and **2b**. Our neutron diffraction measurements also confirm the long-range AFM order at $T_N$ in Fe doped $Sr_2IrO_4$; the data will be published elsewhere. The Fe dopant sharply contrasts other dopants on the Ir site, such as Rh, Mn, Tb [29-32, 35-39] and Co, that readily suppress the AFM order; but it hints certain similarity to Ru doping discussed below [33, 34]. The temperature dependence of $M_a(T)$ and $M_c(T)$ changes significantly below $T_N$. A rapid rise in $M_a(T)$ and $M_c(T)$ below 35 K, T*, whose onset is defined by the valley in M, is likely a result of the polarization of the acceptor states; it gets stronger in magnitude and shifts to higher temperatures with increasing x (**Figs.2a** and **2b**). The upturn of M(T) and the slow increase of $\mu_{eff}$ with x (**Fig.2c**) are an indication that $Fe^{4+}$ is in the intermediate spin state, S=1. The low T magnetic anomaly is not a phase transition because no corresponding anomaly around T* is seen in the heat capacity discussed below. In addition, the magnetic anisotropy between $M_a(T)$ and $M_c(T)$ becomes increasingly weaker as x increases (**Figs.2a** and **2b**), which appears consistent with the shortening of the c axis (**Fig.1d**), thus an enhanced inter-plane magnetic interaction.

Furthermore, the isothermal magnetization M(H) increases considerably with x but becomes less "saturated", compared to that in x=0 (see Inset in **Fig.2b**), suggesting a reduced magnetic canting, which produces the weak ferromagnetic behavior that features x=0. Since the magnetic moment is strongly coupled to the lattice [5, 6, 15, 42, 43], a relaxed Ir-O-Ir bond angle (**Fig.1h**) inevitably weakens the Dzyaloshinskii-Moriya interaction that drives magnetic canting, thus weak ferromagnetism. A change in the magnetic configuration is possible, and is seen in Mn, Rh, or Ru doped $Sr_2IrO_4$ [30, 32, 34].

More data analysis using a Curie-Weiss law retrieves the Curie-Weiss temperature, $\theta_{CW}$, and effective moment, $\mu_{eff}$, which deserves a close examination. $\theta_{CW}$ slowly decreases with x for x < 0.20. For example, $\theta_{CW}$ is merely reduced to 213 K for x = 0.14 from 250 K for x = 0, as shown in a $\Delta\chi^{-1}$ plot in **Fig.2a** (Note that $\chi$ is the magnetic susceptibility; $\Delta\chi = \chi - \chi_o$, where $\chi_o$ is the temperature-independent contribution to $\chi$). $\theta_{CW}$ essentially tracks $T_N$ (**Fig.2c**) as a function of x, dropping precipitously only when x > 0.20. Clearly, the superexchange interaction between magnetic moments supporting the AFM order is surprisingly resilient to Fe doping up to 20%.

This intriguing behavior calls for an understanding. As established above, the $Fe^{4+}$ ion carries four *3d*-electrons with a less than half filled *3d*-shell. Accordingly, the first Hund's rule dictates the total spin to be S = 1 (the $d_{xy}$ is doubly occupied, while the $d_{xz}$ and $d_{yz}$ orbitals have one electron) and the orbital angular momentum is quenched by the crystalline field. Since for x=0 $\mu_{eff} = 0.46$ $\mu_B$/f.u. and the Fe-ions contribute with S=1, the effective moment $\mu_{eff}$ only rises slightly to 0.67 $\mu_B$/f.u. for x=0.14, as shown in **Fig.2c**. Such a small increase in $\mu_{eff}$ for $x \leq 0.14$ implies that the Fe ions are in the S = 1 state, thus contributing a very small amount of magnetic moment to the total value of $\mu_{eff}$. As the Fe doping level further increases, the acceptor bound states start to overlap forming a band which allows the Fe spins to interact with each other giving rise to ferromagnetic correlations. This explains a rapid rise in $\mu_{eff}$ for x > 0.20, where $\mu_{eff}$ increases to 1.90 for x=0.22 $\mu_B$/f.u, and then to 2.73 $\mu_B$/f.u for x=0.31(**Fig.2c**).

It is thus almost certain that the unexpected resilience of the AFM order to the Fe doping for x < 0.20 is primarily because the $Fe^{4+}$ ions have only a S=1 spin, thus carrying essentially small magnetic moments that are too weak to effectively affect the superexchange interaction of the iridate and thus the AFM order. With further increasing Fe doping (x > 0.20), the SOI weakens in the host and the Fe magnetic moments eventually become strong enough to suppress $T_N$ when x > 0.2. That $T_N$ and $\mu_{eff}$ closely track each other in an opposite fashion with increasing x reinforces this argument, as evidenced in **Fig. 2c**. A strikingly similar trend is also reported in our early study of $Sr_2Ir_{1-x}Ru_xO_4$ in which the doping of $Ru^{4+}$(*4d⁴*) ions offer four *4d*-electrons and a S=1 spin state [33, Fig.5].

*$Sr_2Ir_{1-x}Co_xO_4$.* The $Co^{4+}$ ion carries five *3d*-electrons, thus, assuming the high spin state (one electron per orbital) Hund's rules yield S = 5/2 with the orbital angular momentum being completely quenched. It is therefore not surprising that, unlike Fe doping, Co doping readily suppresses the native AFM order, as shown in **Fig.2d-2e**. The Curie-Weiss temperature $\theta_{CW}$ decreases drastically from 250 K for x=0 to 32 K for x=0.11 (see $\Delta\chi^{-1}$vs T in **Fig.2d**). A magnetic peak briefly occurs in a narrow temperature range of 8 K - 10 K for $0.08 \leq x < 0.14$, vanishing at x > 0.14; it could be either the fading native AFM order or an emergent new magnetic state originating from the interaction between the Co spins. The magnetic state has a vanishing magnetic anisotropy (Inset in **Fig.2e**), implying a weakening SOI effect. Importantly, the effective moment $\mu_{eff}$ rises at a fast rate and is accompanied by an equally fast vanishing $T_N$ (**Fig. 2f**). The closely locked opposite trends of $\mu_{eff}$ and $T_N$ as a function of x in $Sr_2Ir_{1-x}Co_xO_4$ further strengthens the key notion that it is the value of S of the substituting ion for the Ir ion that predominately determines the magnetic properties in doped $Sr_2IrO_4$.

### (c) Electrical Resistivity

Despite the vastly different magnetic properties between $Sr_2Ir_{1-x}Fe_xO_4$ and $Sr_2Ir_{1-x}Co_xO_4$, the transport properties of these compounds exhibit a great deal of similarity, highlighting a near irrelevance of the magnetic state to the electronic state in $Sr_2IrO_4$.

As shown in **Figs. 3a-3b**, either Fe or Co doping almost instantly delocalizes electrons, drastically reducing the resistivity, ρ, by up to *eight* orders of magnitude for the a-axis resistivity $\rho_a$ and *six* orders of magnitude for the c-axis resistivity $\rho_c$. For $Sr_2Ir_{1-x}Fe_xO_4$ the conducting behavior arises from the impurity band generated from the overlap of the extended acceptor bound states, while for $Sr_2Ir_{1-x}Co_xO_4$ the impurity band is due to the percolation of the bound states, a consequence of the scattering off the isoelectronic $Co^{4+}$ impurities. The size of the bound states is of the order of one unit-cell, but the percolation limit depends on the neighboring bonds (first, second, third) that have to be considered [40]. It can be calculated straightforwardly for a Bethe lattice of coordination z, $x_{cr} = [z(z-1)]^{-1}$, i.e. for motion in the ab plane z = 4 and $x_{cr} \sim 8\%$ and for the 3D crystal z = 6 and $x_{cr} \sim 3\%$. The percolation threshold for Fe is much less than for Co.

It is usual to call "metallic" behavior for positive dρ/dT and "insulating" behavior for negative slope. In $Sr_2Ir_{1-x}Co_xO_4$, the metallic state exists in a wide temperature range, as shown in **Fig.3d**. The metallic behavior emerges in $\rho_a$ for x as small as 0.08 in both $Sr_2Ir_{1-x}Fe_xO_4$ and $Sr_2Ir_{1-x}Co_xO_4$, where $\rho_a$ drops from $10^4$ Ω cm for x=0 to $10^{-4}$ Ω cm (see **Figs.3c-3d**). Note that some resistivity data are collected up to 700 K. In fact, the insulating behavior already vanishes above 100 K at mere x=0.02 of Fe doping, in which $d\rho_a/dT \sim 0$ (**Fig.3c**). However, the c-axis resistivity $\rho_c$ for both Fe and Co doping still exhibits a negative slope of $d\rho_c/dT$ despite the drastically reduced magnitude of $\rho_c$ (**Fig.3b**). The contrasting behavior between $\rho_a$ and $\rho_c$ suggests a highly anisotropic metallic state in the materials discussed below.

For $Sr_2Ir_{1-x}Fe_xO_4$, a pronounced upturn in $\rho_a$ and $\rho_c$ below 35 K is closely associated with the magnetic anomaly T* (**Fig.3e**) resulting from the spin polarization of the electrons localized in the acceptor band. With increasing Fe doping the interaction between Fe spins becomes larger and so does T* (**Figs.2a-2b**). For N Fe impurities the donor band accommodates up to 2N electrons. If the band is spin-polarized the up-spin subband is filled and the down-spin subband empty, i.e. the system is insulating. The recovered insulating behavior for x = 0.31 (**Fig.3b**) is attributed to (weak) localization due to scattering off randomly distributed Fe ions (recall that the isostructural $Sr_2FeO_4$ is an AFM insulator with a Néel temperature at 60 K [44]). It is interesting to note that the magnetic and transport data of $Sr_2Ir_{1-x}Fe_xO_4$ indicate that the metallic state exists only in a region where the long-range AFM order persists and gives way to a more insulating state when the AFM order vanishes (e.g., x=0.31), hinting unconventional correlation between magnetic and electronic states.

In $Sr_2Ir_{1-x}Co_xO_4$, the metallic state exists in a wide temperature range, as shown in **Fig.3d**. Most remarkably, $\rho_a$ exhibits a pronounced linearity with temperature up to 700 K, with no sign of saturation. Note that for the iridates, the Debye temperature, $\theta_D$, ranges from 390 to 440 K. The linear-temperature dependence of resistivity at high temperatures is a hallmark of many interesting "bad" metals, such as superconducting cuprates [45] and, more recently, twisted bilayer graphene [46]. High-temperature resistivity of conventional metals, which is driven by strong electron-phonon scattering, saturates at sufficiently high temperatures higher than $\theta_D$ when the mean free path of a quasiparticle becomes shorter than the de Broglie wavelength, according to the Boltzmann transport theory. The absence of resistivity saturation suggests a breakdown of conventional theoretical models, and the prevailing view is that this phenomenon is associated with collective fluctuations, possibly spin fluctuations, but the physics of it has remained one of time-honored intellectual challenges [47].

The metallic state is confined to the basal plane. The interplane or c-axis resistivity $\rho_c$ remains having a negative slope of $d\rho_c/dT$ (**Fig.3b**) for both systems. The anisotropy between $\rho_c$ and $\rho_a$ is large; for example, the ratio $\rho_c/\rho_a \approx 100$ for x=0.18 of Co doping (see **Fig.3f**). Similar behavior was observed in the cuprate superconductors, anisotropic organic conductors and, more recently, in compressed $Sr_3Ir_2O_7$ [48]. In essence, the anisotropic metallic behavior is recognized as a result of confined coherence [49, 50]. For the perovskite systems such as the cuprates and the iridates, the single electron coherence is confined to the basal planes so that an inference effect or coherent electron hopping between the basal planes becomes unlikely, thus resulting in *incoherent* non-metallic transport along the c axis. A confined metallic state can occur only in strongly anisotropic and correlated non-Fermi liquids [50].

### (d) Specific heat

To evaluate the electronic correlations, we now examine the specific heat, C(T), culled in 0.05 K < T <50 K for both $Sr_2Ir_{1-x}Fe_xO_4$ and $Sr_2Ir_{1-x}Co_xO_4$. As shown in **Fig.4a**, C(T) for both x=0.11 of Co doping and x=0.14 of Fe doping is not only different but also considerably larger than that for x=0, corroborating the changed ground state. For x=0.11 of Co doping, a weak but visible hump near 10 K in C(T) is attributed to the magnetic order seen in the magnetization (**Figs.2d-2e**). On the other hand, C(T) for x=0.14 of Fe doping exhibits no anomaly at T* near 30

K, but rapidly decreases below 2.5 K, becoming vanishingly small at 0.05 K. The Sommerfeld coefficient, $\gamma$, estimated from a plot of $C(T)/T$ vs $T^2$ is approximately 30 mJ/mole K$^2$ for x=0.14 of Fe doping and 19 mJ/mole K$^2$ for x=0.11 of Co doping, compared to 4 mJ/mole K$^2$ for x=0 (**Fig.4b**; note that $\gamma$ is extrapolated from the data above the magnetic anomaly marked by the blue and red dashed lines). The large values of $\gamma$ for the doped iridates reflect the significant electronic contribution of the emergent metallic state to C(T). Based on the values of $\gamma$ and the lattice parameters, the effective mass, $m_{eff}$, is estimated to be 10.2 $m_e$ and 5.9 $m_e$ for the Fe and Co doped samples, respectively ($m_e$ is electron rest mass). Furthermore, the Wilson ratio, $R_w \sim \chi_o/\gamma$, is 2.7 and 2.6, where $\chi_o$ = 1.2 x10$^{-3}$ emu/mole and 7.0 x 10$^{-4}$ emu/mole for the Fe and Co doped samples, respectively. The significantly enhanced values of both $m_{eff}$ and $R_w$ strongly indicate that electrons are not only highly correlated in Sr$_2$Ir$_{1-x}$**Fe**$_x$O$_4$ and Sr$_2$Ir$_{1-x}$**Co**$_x$O$_4$, but also non-locally correlated, i.e. the impurities interact with each other.

**(e) Hall effect**

The charge-carrier density, n, obtained from Hall effect measurements also indicates a consistent, drastic increase for the Fe and Co doped samples. As shown in **Fig.4c**, the highest absolute values of n at high temperatures increases from 10$^{26}$/m$^3$ for x=0 to 10$^{27}$/m$^3$ and 10$^{28}$/m$^3$ for x=0.09 of Fe doping and x=0.11 of Co doping, respectively, and these values of n are comparable to those of metals, a consequence of the impurity bands. Note that n for Co doped Sr$_2$IrO$_4$ is negative, indicating that charge carriers are primarily electrons. Furthermore, the temperature dependence of n for the three compounds is both different and telling. For x=0, n increasing with temperature, suggests thermal activation, typical of an insulator or semiconductor. The value of n for x=0.09 of Fe doping is one order of magnitude higher than that for x=0 but its temperature dependence also suggests a significant role of thermal activation; thus, the transport properties are likely governed by both the added holes to the acceptor band and their thermal activation. On the other hand, for x=0.11 of Co doping, the absolute value of n rises initially and then approaches a near saturation above 50 K. This clearly indicates that the transport properties are driven by itinerant electrons in the percolated impurity band, rather than thermally activated.

## IV. Conclusions

The observations presented in this study provide insights into the iridates. In Fe doped $Sr_2IrO_4$, a metallic state emerges from the acceptor band as the native AFM order persists -- there is no sign of a correlation between the native AFM and insulating states (**Fig.5a**). This behavior sharply contrasts that in Co doped $Sr_2IrO_4$ in which the transport and magnetic properties closely track with each other in a fashion commonly seen in other correlated systems (**Fig.5b**). For the Co doped $Sr_2IrO_4$ the properties arise from the impurity without acceptors or donors. The contrasting behavior suggests that the origin of the AFM-Mott state has no close association with the electronic conduction state. The absence of a correlation between the AFM and Mott in doped $Sr_2IrO_4$ sharply contrasts that of other correlated materials whose signature is often a strong AFM and Mott correlation.

This study concludes that the distinctly different response of the magnetic state to $Fe^{4+}$ (*3d⁴*) and $Co^{4+}$ (*3d⁵*) doping is a consequence of the different spin state of the substituting ion for the Ir ion. The impurity bands arising from the percolation of the bound states yield a different metallic behavior in the two systems due to the difference between acceptor states and charge neutral substitutions. The different electronic configurations of $Ir^{4+}$, $Fe^{4+}$ and $Co^{4+}$ are shown in **Figs.5c-5e**.

This insight also provides a universal explanation for changes of $T_N$ due to *d*- or *f*-electron doping for Ir *5d*-electrons in $Sr_2IrO_4$. The emergent metallic state in Fe or Co doped $Sr_2IrO_4$ is assisted by the significantly reduced octahedral IrO$_6$ rotation, which facilitates electron hoping in the basal plane. It is peculiar that the basal plane network of IrO$_6$, which facilitates coherent electron hopping, is resilient to the high level of doping. These observations may provide a new pathway to discoveries of new states in the iridates.

**Acknowledgement:** This work is supported by NSF via grant DMR 1903888. GC is thankful for useful discussions with Peter Riseborough and Itamar Kimchi.

**Captions**

**Fig.1. Structural properties of $Sr_2Ir_{1-x}Fe_xO_4$ and $Sr_2Ir_{1-x}Co_xO_4$ at 100 K: (a)** The unit cell of $Sr_2IrO_4$; **(b)** the basal plane of $Sr_2IrO_4$. The x-dependence of **(c)** the a axis, **(d)** the c axis, **(e)** the ratio of c/a, **(f)** the unit cell volume V, **(g)** the bond length Ir-O2 and **(h)** the bond angle Ir-O-Ir.

**Fig.2. Magnetic properties of $Sr_2Ir_{1-x}Fe_xO_4$ and $Sr_2Ir_{1-x}Co_xO_4$:** Temperature dependence of **(a)** the a-axis magnetization $M_a$ and **(b)** the c-axis magnetization $M_c$ for $Sr_2Ir_{1-x}Fe_xO_4$; **(c)** the Néel temperature $T_N$ (left scale) and the effective moment $\mu_{eff}$ (right scale) as a function of Fe doping x. Temperature dependence of **(d)** the a-axis magnetization $M_a$ and **(e)** the c-axis magnetization $M_c$ for $Sr_2Ir_{1-x}Co_xO_4$; **(f)** the Néel temperature $T_N$ (left scale) and the effective moment $\mu_{eff}$ (right scale) as a function of Co doping x. Note that the dashed line in (b) is a guide for the eye, indicating the shifting T*.

**Fig.3. Transport properties of $Sr_2Ir_{1-x}Fe_xO_4$ and $Sr_2Ir_{1-x}Co_xO_4$:** Temperature dependence of the electrical resistivity $\rho$ for x = 0 (black), and representative doping x=0.11 of Co (red) and 0.14 of Fe (blue) for **(a)** the a-axis resistivity $\rho_a$ and **(b)** the c-axis resistivity $\rho_c$. Note that the reduction of $\rho_a$ and $\rho_c$ is as much as eight and six orders of magnitude, respectively. Temperature dependence of the a-axis resistivity $\rho_a$ for **(c)** $Sr_2Ir_{1-x}Fe_xO_4$ and **(d)** $Sr_2Ir_{1-x}Co_xO_4$. Note $\rho_a$ is as low as $10^{-4}$ $\Omega$ cm for x=0.07 (Fe) and 0.08 $\Omega$ cm (Co). **(e)** Temperature dependence of the a-axis resistivity $\rho_a$ (left scale) and magnetization $M_a$ (right scale) for x=0.14 of Fe doping and **(f)** the a-axis resistivity $\rho_a$ and the c-axis resistivity $\rho_c$ for x=0.18 of Co doping.

**Fig.4. Specific heat and charge-carrier density of $Sr_2Ir_{1-x}Fe_xO_4$ and $Sr_2Ir_{1-x}Co_xO_4$: (a)** Temperature dependence of specific heat C and **(b)** C/T vs $T^2$ for x = 0 (black), 0.11 of Co doping (red) and 0.14 of Fe doping (blue). **(c)** Temperature dependence of the charge-carrier density n for x = 0 (black), 0.09 of Fe doping (blue) and 0.11 of Co doping (red).

**Fig.5. Phase diagrams** for **(a)** $Sr_2Ir_{1-x}Fe_xO_4$ and **(b)** $Sr_2Ir_{1-x}Co_xO_4$ generated based on the results, where M = confined metallic state, I = insulating state and PM = paramagnetism. **(c)** Schematics of the crystalline field and spin orbit interaction splitting for the **$Ir^{4+}$**($5d^5$) ions. Schematics of **(d)** the high spin state (HS) S=5/2 of **$Co^{4+}$**($3d^5$) impurities and (e) the intermediate spin state (IS) S=1 of **$Fe^{4+}$**($3d^4$) acceptor impurities.

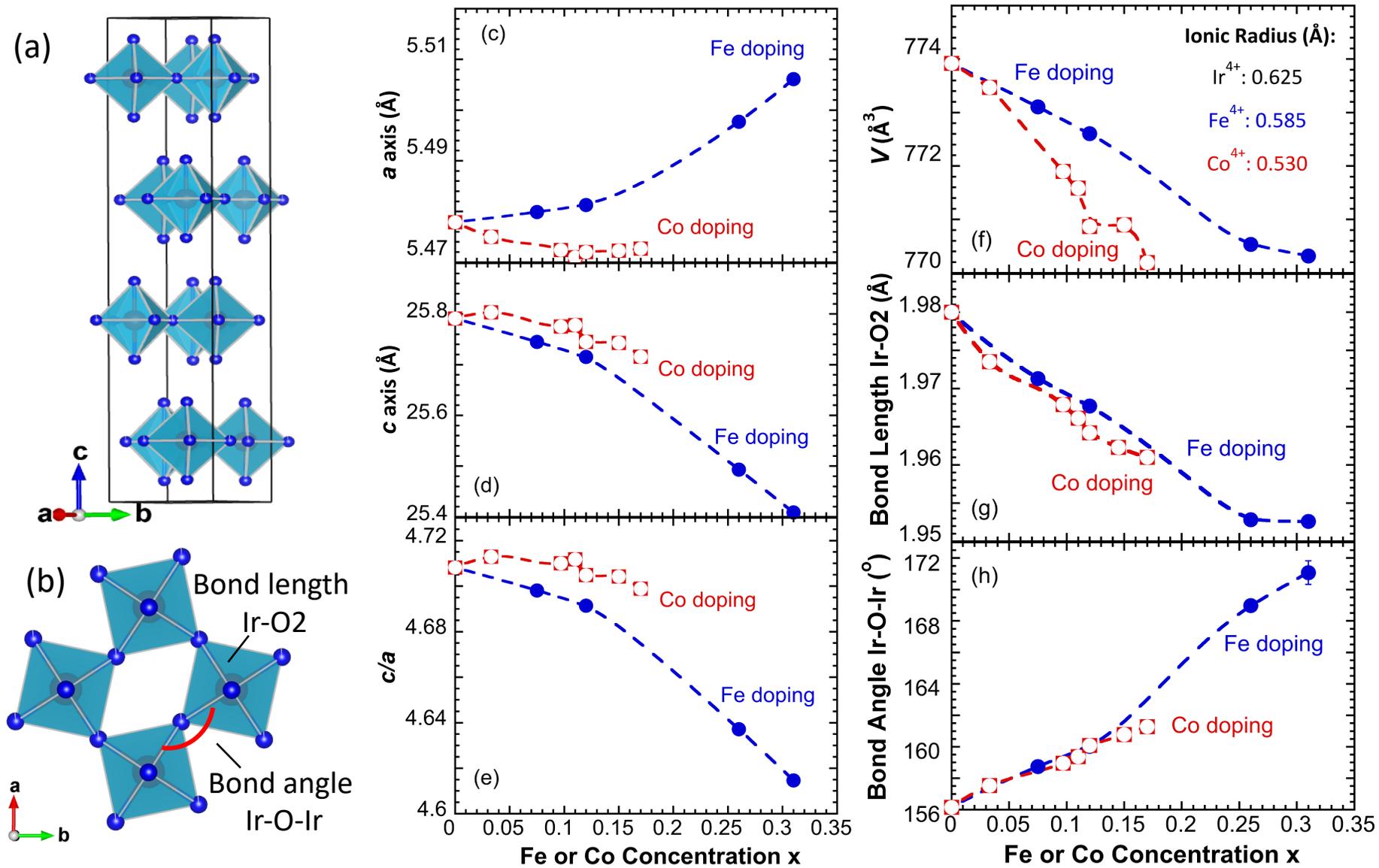

Figure 1

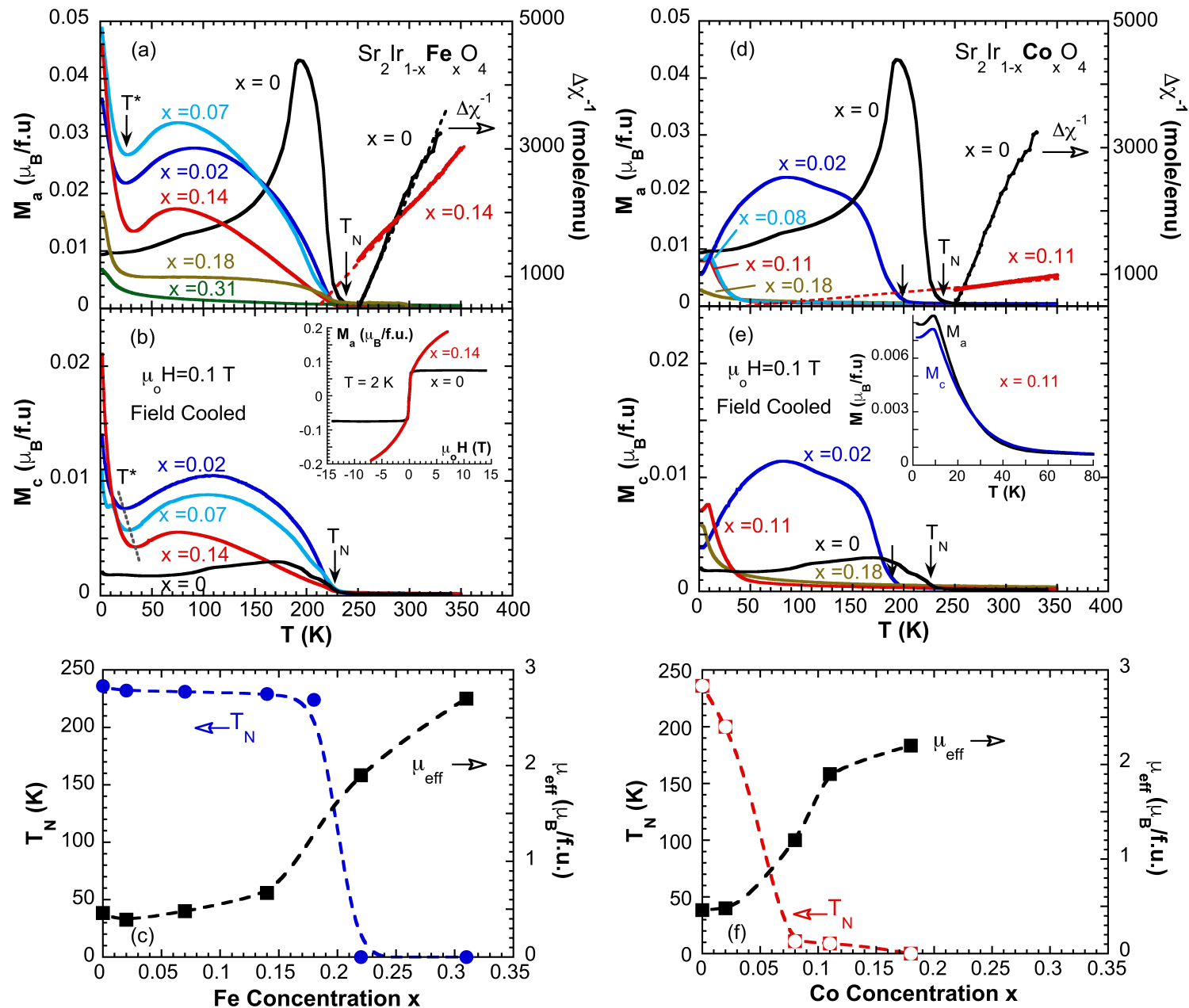

Figure 2

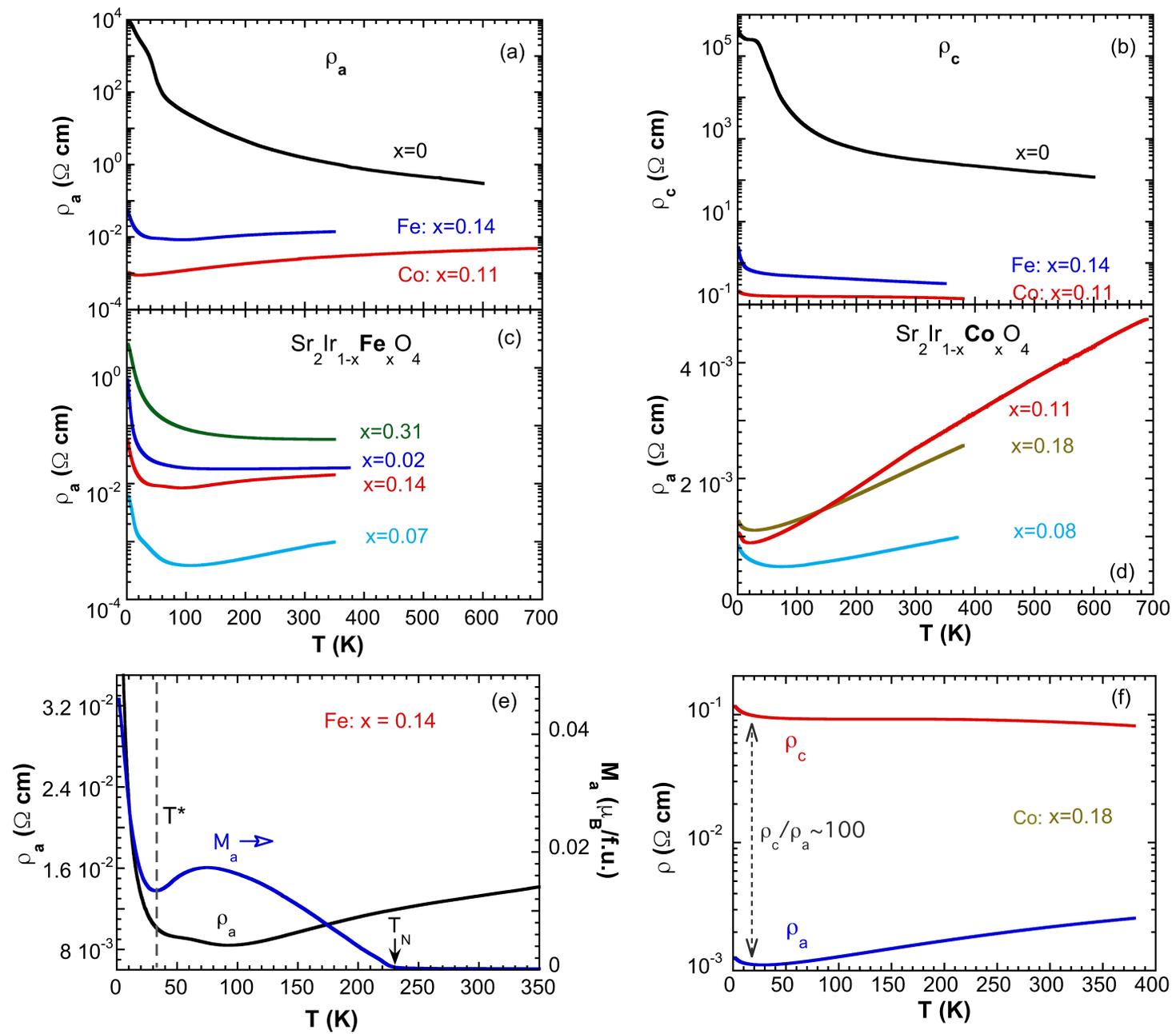

Figure 3

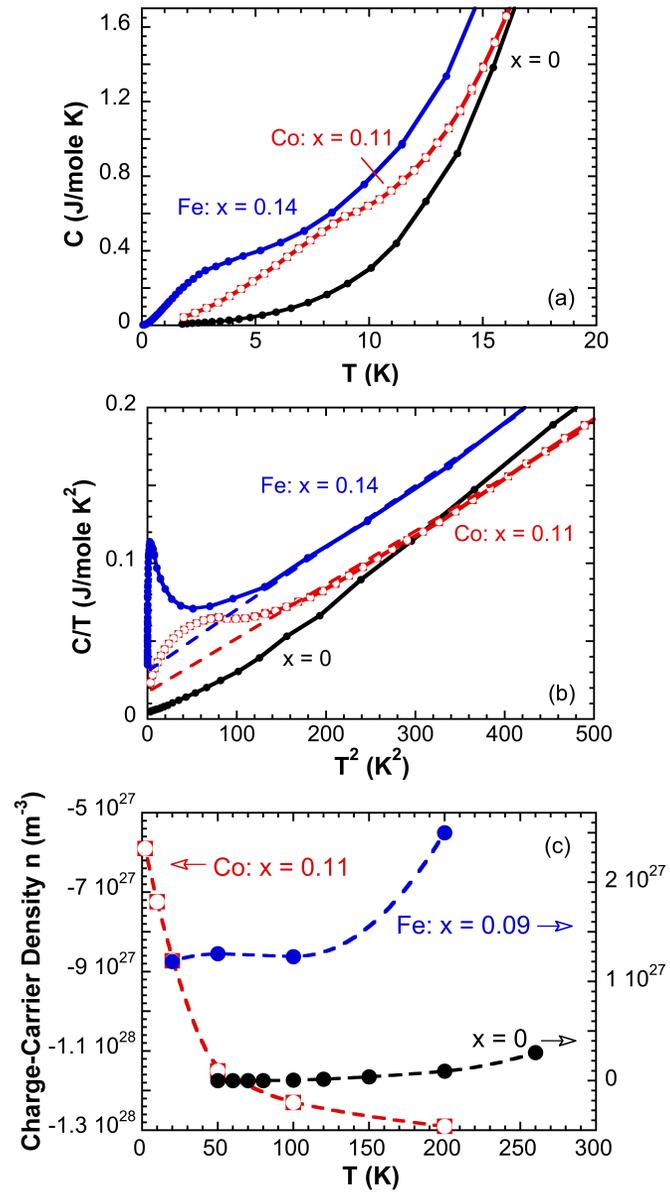

Figure 4

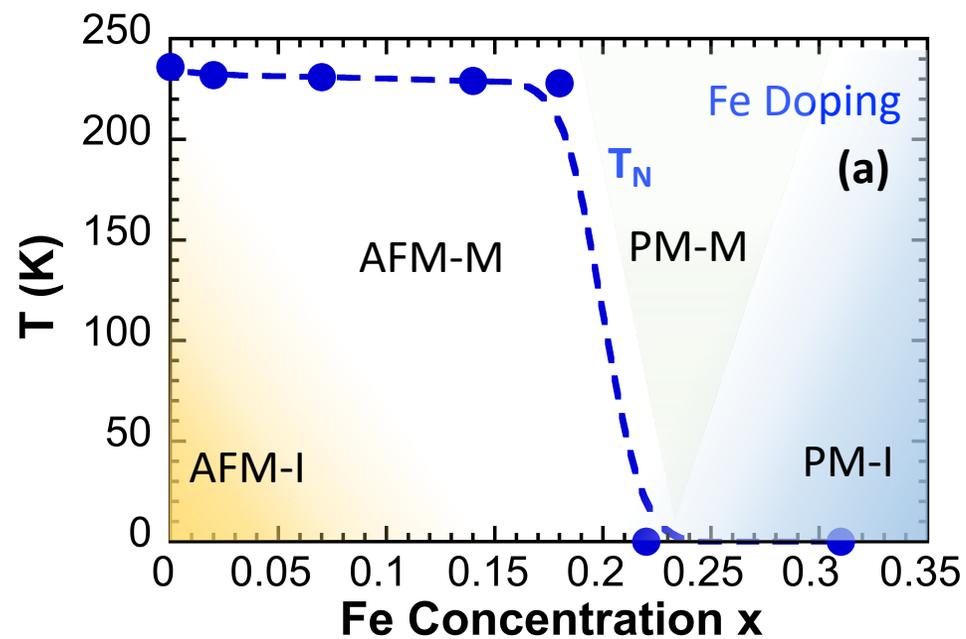
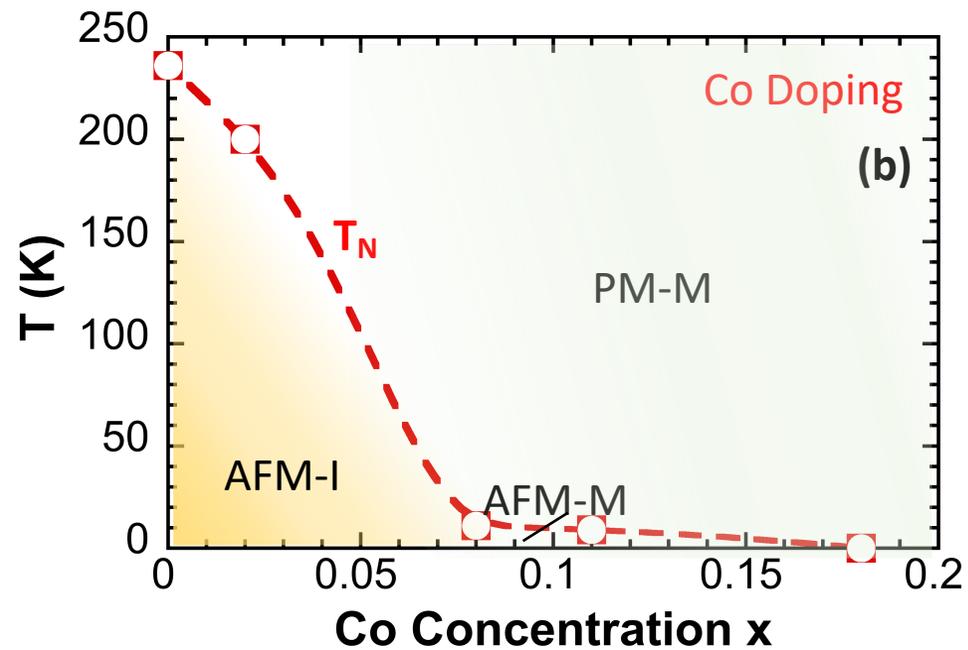
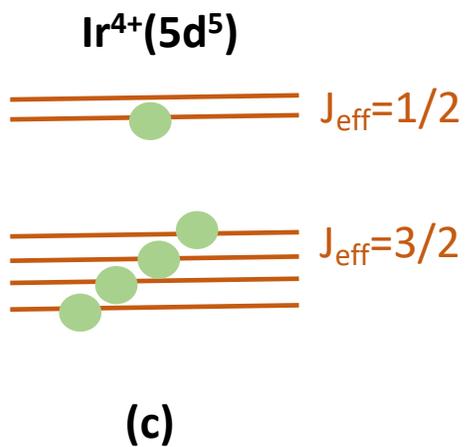
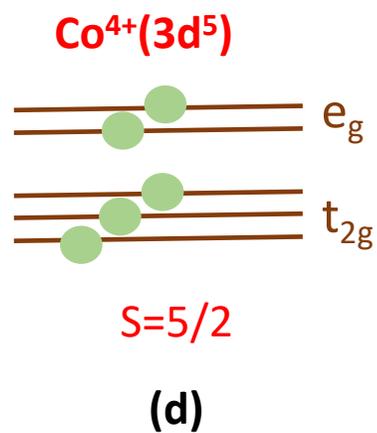
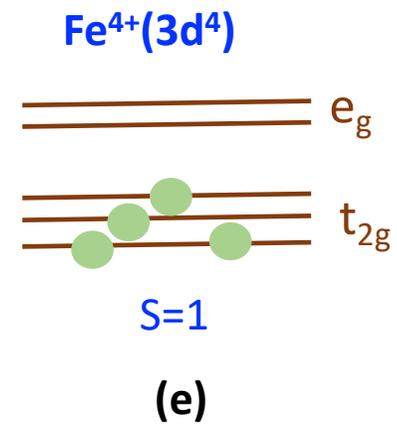

Figure 5